# Effects of Free-ranging Livestock on Sympatric Herbivores at Fine Spatiotemporal Scales

**RONGNA FENG,** [1] **XINYUE LÜ,** [1] **TANMING WANG,** [2] **JIAWEI FENG, YIFEI SUN, YU GUAN, LIMIN FENG, JIANPING GE,** *Ministry of Education Key Laboratory for Biodiversity Science and Engineering & College of Life Sciences, Beijing Normal University, Beijing, 100875, China.*

**WENHONG XIAO,** *Institute of Zoology, Chinese Academy of Sciences, Beijing 100101, China.*

**JAMES L. D. SMITH,** *Department of Fisheries, Wildlife and Conservation Biology, University of Minnesota, St Paul, MN 55108, USA*

[1] *These authors contributed equally to this work*

[2] *E-mail:* wangtianming@bnu.edu.cn (https://orcid.org/0000-0003-3370-0209)

## ABSTRACT

Understanding wildlife-livestock interactions is crucial for the design and management of protected areas. Free-ranging livestock degrade the food and habitat of the endangered Amur tiger and Amur leopard in a newly established national park in Northeast China, but quantitative assessments of how livestock affect the use of habitat by the major ungulate prey of these predators are very limited. Here, we examine livestock-ungulate interactions using data from a large-scale camera-trap study. We used N-mixture models, two-species occupancy models and activity pattern overlap to understand the effects of cattle grazing on three ungulate species (wild boar, Siberian roe deer and sika deer) at a fine spatiotemporal scale. Our results showed that including the biotic interactions with cattle had significant negative effects on encounters with three ungulates. In particular, sika deer were displaced as more cattle encroached on forest habitat, inferred by the low levels of co-occurrence with cattle and limited spatiotemporal overlap. These insights can help to refine strategies for conserving tigers, leopards and their natural prey in human-dominated transboundary landscapes. Accounting for the impact of cattle on biodiversity while simultaneously addressing the economic needs of local communities should be key priority actions for natural resource managers.

Livestock production covers nearly a quarter of the land surface of the planet and presents a substantial threat to native fauna by competing for limited space and resources (Robinson et al. 2014, Schieltz and Rubenstein 2016, Zhang et al. 2017). Livestock grazing can greatly intensify pressure on local wildlife, as it alters their temporal activity patterns and decreases their spatial distribution, habitat use and food availability (Madhusudan 2004, Herfindal et al. 2017, Pudyatmoko 2017, Valls-Fox et al. 2018). As the human population increases globally, livestock encroachment into protected areas has generated unintended consequences. For instance, Pudyatmoko (2017) found that large carnivores and herbivores were absent in areas with livestock in Indonesia, and some ungulate species even altered their activities from diurnal to nocturnal in the presence of livestock. A recent study also showed that livestock grazing had significant negative impacts on the occupancy of carnivores and their ungulate prey in protected areas in the Hyrcanian forests in Iran (Soofi et al. 2018). Similar results were observed in a wildlife reserve in Southwest China, where giant pandas (*Ailuropoda melanoleuca*) and sympatric species were displaced as livestock encroached on forest habitat (Li et al. 2017a, Zhang et al. 2017). These findings raise great concern about wildlife management and conservation policies in landscapes that are increasingly dominated by disturbance from cattle grazing.

Through the negative effects on large herbivores, livestock grazing causes the loss and degradation of habitats for large carnivore species. In the temperate mixed-forests of northeast Asia, sympatric carnivore species such as the endangered Amur tiger (*Panthera tigris altaica*) and Amur leopard (*Panthera pardus orientalis*) face threats from uncontrolled cattle grazing (Wang et al. 2016). Despite substantial recovery efforts, these iconic carnivores are still mainly confined to a narrow strip of land along the northeast border with Russia (Feng et al. 2017). Amur tigers and leopards also prey on livestock, which causes further conflicts with humans in addition to habitat loss, poaching, prey depletion and disease (Gilbert et al. 2015, Miquelle et al. 2010, Tian et al. 2011). Past efforts to reduce deforestation, especially "The Natural Forests Protection Program" initiated in 1998, while simultaneously encouraged local people to raise cattle that then freely ranged in forests in Northeast China. Cattle compete with wild ungulates, potentially reducing the availability of the natural prey of carnivores, and the combination of cattle grazing and other human activities is restricting the expansion of tigers and leopards further into China (Wang et al. 2016, Wang et al. 2017, Wang et al. 2018). As an alternative to logging, livestock production has become the most prevalent human disturbance and is a main driver of biodiversity loss in Northeast China (Machovina et al. 2015).

In seeking to create a protected area for tigers and leopards, the Chinese government recently initiated a large national park along the China-Russia border. They plan to shift forest management away from livestock grazing to create suitable habitat for declining populations of tigers, leopards and their wild prey, while also providing important ecological services to support human livelihoods (McLaughlin 2016). It is important that conservation initiatives target the recovery of major ungulate prey species, but to date, there have been no quantitative assessments of how livestock affect the abundance and distribution of major ungulate prey of Amur tigers and leopards at large scales. Understanding these effects has become one of the most important research needs to inform the design and management of this newly established national park (McLaughlin 2016).

Here, we present a fine-scale (i.e. camera locations) analysis of the spatiotemporal use patterns of large wild ungulates in response to cattle grazing in Northeast China using data from a large-scale camera-trap study. We focus on three main ungulate prey species (sika deer *Cervus nippon*, wild boars and Siberian roe deer *Capreolus pygargus*) that collectively account for 92% of the tiger diet (21%, 49% and 22%, respectively) and 87% of the leopard diet (50%, 18% and 19%, respectively) in winter (Sugimoto et al. 2016). We investigated the livestock-ungulate interactions in combination with environmental factors using N-mixture and co-occurrence models that account for imperfect detection (MacKenzie et al. 2004, Royle 2004). We hypothesized that all wild ungulates would demonstrate a lack of co-occurrence with cattle and exhibit fine-scale avoidance behaviour. Specifically, we predicted that sika deer may show lower tolerance of livestock disturbance than wild boar and Siberian roe deer (hereafter roe deer), which are known to be generalist species. Finally, we discuss the management actions required to address declines of large ungulates in the temperate forest landscape in Northeast China.

**STUDY AREA**

This research was conducted in the northern portion of the Changbai Mountains in Jilin Province, China, adjacent to southwestern Primorsky Krai, Russia, to the east and North Korea to the southwest (Fig. 1). The approximately 5000-km$^2$ study area forms the core of a potential recovery landscape for tigers and leopards in a new national park in China (Hebblewhite et al. 2012, Wang et al. 2016). Elevations range from 5 to 1477 m. The climate is characterized as a temperate continental monsoon with an average annual temperature of 5.60 °C (± 1.30 °C) and a frost-free period of 110–160 days/yr. The annual average precipitation was 618 mm (± 68 mm) during 1990–2010, with the most precipitation occurring in the summer from June to August. Forest cover is more than 92% and the majority of forests have been converted into secondary deciduous forests over the past 5 decades (Li et al. 2009).

Free-range cattle grazing is one of the main economic activities in the study area, with cattle densities ranging from 8 to 12 livestock per km$^2$ (Wang et al. 2016, Li et al. 2017b). Cattle density is *ca*. 4–6 times the density of the three wild prey (wild boars 0.53 individuals/km$^2$, roe deer 1.22 and sika deer 0.15, respectively) (Qi et al. 2015), with especially high densities of cattle grazing in forests from April to October. When food was scarce in winter, herds were moved to the village. According to our field survey of a total area of *ca*. 3500 km$^2$ in Hunchun, there are at least 280 family-based continuous ranches that cover an area of more than 1200 km$^2$. In addition, local ranchers have built an intricate network of wire or electric fences in the forest to manage their cattle, which has the potential to negatively affect wildlife through direct mortality and reducing landscape connectivity (Harrington and Conover 2006, Gadd 2012). Cattle weight more than 300 kg, which much greater than three wild ungulates (wild boar, roe deer and sika deer) with body size ranging 30-110 kg (Dou et al. 2019). Cattle can reduce the plant biomass in the shrub-herb layer by ~ 24% in our study area (Wang et al. 2019).

## METHODS

### Camera Trap Survey

Camera trapping was conducted continuously from August 2013 to July 2014 (Fig. S1). We established 3.6 × 3.6 km grids to guide the placement of 356 camera trap stations throughout the study area. We deployed at least one camera per grid cell and excluded any non-forest habitat within all grids; on average, adjacent camera stations were 2.36 km apart. We maximized detection probability by placing cameras at sites where tigers, leopards, and their prey were likely to travel (e.g., along ridges, valley bottoms, trails, forest roads and near scent-marked trees). We deployed cameras (LTL 6210M, Shenzhen, China) along forest roads ($n$ =199 sites) and game trails ($n$ =157 sites), where they were fastened to trees approximately 40–80 cm above the ground and programmed to take photographs 24 h/day with a 1-min interval between consecutive events. We visited each camera 3–5 times a year to download photos and check the batteries.

**Covariates**

We considered a suite of abiotic and biotic covariates that could potentially influence the spatial distribution of the three ungulates in this area (Table 1). Specifically, we considered elevation, topographic position index (TPI; e.g., finer-scale depressions or ridges) (De Reu et al. 2013), percent tree cover (PTC), the nearest distance to the Russian border, cattle encounter rates and anthropogenic activity. We also tested for a quadratic effect of elevation and TPI assuming the three ungulates prefer intermediate levels of these covariates. TPI was calculated using a circular neighbourhood with a 1-km radius from the Shuttle Radar Topography Mission (SRTM) 30-m digital elevation model. The PTC was derived for each camera station from 250-m Moderate Resolution Imaging Spectroradiometer (MODIS) imagery (product MOD44B) of the study area. We used distance to the Russian border as a measure of the effect of the source wildlife populations on occurrence. For the spatial measures of anthropogenic activity, we calculated the nearest distance to settlements and roads as well as encounter rates of humans (i.e., people on foot) and vehicles. All distance covariates were calculated in ArcMap 10.1 for each camera station.

To understand species interactions, we considered the cattle encounter rate (i.e., a quantitative measure of grazing intensity) or presence/absence (0/1) as predictors of the activity of the three ungulates. Given that sika deer outcompete roe deer (Aramilev 2009), we added sika deer to the roe deer models. We analysed tigers, leopards, wild boar, sika deer, roe deer, cattle, and humans and vehicles as camera trap "entities" and calculated encounter rates for each entity at each camera-trap station as the number of detections per 100 camera-trap days using a 30-min period of independence per entity for the entire sampling period (O'Brien et al. 2003).

The covariates used to model the baseline detection probability included camera days (total number of days that each camera was in operation) as a measure of effort and two predation risk factors (tiger and leopard activity). We also allowed for detection probabilities to vary by trapping occasions.

We tested for collinearity among covariates using two methods. First, a variance inflation factor (VIF), which measures multicollinearity among variables, was calculated for all covariates and those with a VIF < 3 were retained in the model. Then, Pearson's correlation coefficients (r) were calculated to further exclude highly correlated variables with a $|r| > 0.7$ (Dormann et al. 2013). All continuous covariates were scaled to a mean of 0 and a standard deviation of 1 prior to the analyses.

**Habitat Use Modelling**

We used N-mixture models (Royle 2004) to assess the relative effects of abiotic and biotic covariates on the spatial use of the three ungulates at each camera site using 12 months of camera trap data. Since camera trap designs are considered "plotless designs", N-mixture models have been used to estimate activity rates from spatially and temporally replicated counts of unmarked animals while accounting for imperfect detection (Shamoon et al. 2017b). Thus, for each camera site, we used month as the temporal sampling unit (i.e., survey occasion) and accumulated encounters within each occasion as counts (Xiao et al. 2018). The number of encounters, which was considered a measure of activity rate, indicated whether a site was more or less likely to be visited by animals (Rowcliffe et al. 2008, Shamoon et al. 2017a), so we also used the encounter index as an indicator of the intensity of habitat use by the three ungulates based on the assumption that habitat conditions are directly related to the number of times that a location is visited by the target species (Boyce and McDonald 1999). For example, if an animal forages or shelters in an area, it will be photographed for longer periods of time and thus, have higher encounter rates.

We began our modelling procedure by using the monthly encounter rates at each camera trap to build N-mixture models for each species with all covariates (hereafter, ENV model). In this step, we removed covariates that were not significant for any species using a stepwise selection procedure, and we then established two additional models for each wild ungulate species. We added cattle encounter rate information (hereafter, cattle.num model) and the presence or absence of cattle (hereafter cattle.pres model) into the first model to test levels of tolerance to cattle disturbance. For each ungulate, we determined the most supported model using AIC (Burnham and Anderson 2012). We used computationally efficient graphical checks and overdispersion measures to assess the goodness-of-fit of N-mixture models using the R package `nmixgof` (Knape et al. 2018). The coefficient estimates of the final model were considered significant if their 95% CIs did not include zero. We used a zero-inflated Poisson variant of the N-mixture since there were many zeroes in the data, which provided a better fit to the data compared to a Poisson variant as determined by a likelihood ratio test (LRT) (Zuur et al. 2012). There is evidence that the zero-inflated Poisson variant of N-mixture model is significantly superior to the Poisson variant (LRT: L = 100.92, $p < 0.001$ for wild boar; L = 83.00, $p < 0.001$ for roe deer; L = 61.07, $p < 0.001$ for sika deer; L = 275.02, $p < 0.001$ for cattle). All N-mixture models were fit using the R package `unmarked` (Fiske and Chandler 2011). Deer and wild boar in our study area are highly sedentary, with no detectable seasonal movement patterns (Hojnowski et al. 2012), we thus reported the result of N-mixture models using 12 month data which reflecting the influences of long-term livestock husbandry practices on the intensity of habitat use by the three ungulates.

**Spatial Co-occurrence**

We investigated the potential co-occurrence between the dominant cattle (A) and the subordinate three ungulates (B) by fitting two-species habitat occupancy models to the camera-trapping data from the study area. We used conditional parameterization to estimate each parameter (Table S1) (Richmond et al. 2010) and assumed that the occupancy and detection of the three ungulates were dependent on the presence or absence of the cattle. As we assessed fine-scale space use, we interpreted occupancy as the probability of the use of a camera site. We aggregated 2-week survey periods into a single sampling occasion and constructed detection histories for cattle and the three ungulates for each camera site, resulting in 26 temporal replicates.

We estimated a species interaction factor (henceforth, *SIF*) for each species combination and considered *SIF* < 1.0 to be evidence of apparent spatial segregation, *SIF* > 1.0 to be apparent spatial overlap, and *SIF* =1.0 to be evidence of site-use independence. We implemented the model in the programme PRESENCE 11.8 (Hines 2017).

**Daily Activity Patterns**

All detection events were used to create 24-h activity patterns by ignoring the calendar date for each entity. We used kernel density estimation and trigonometric sum distribution to estimate the probability density functions of the activity patterns (i.e., density of activity) for each entity. We only recorded 10 detections of cattle in the winter. Thus, we estimated the overlap coefficient ($\Delta$) using the R package `overlap` (Ridout and Linkie 2009) to assess activity pattern overlap between cattle and each wild ungulate species from April to October (Ridout and Linkie 2009, Meredith and Ridout 2017). The coefficient ranges from 0 (no overlap) to 1 (complete overlap) with a low degree of overlap indicating temporal avoidance. We obtained 95% confidence intervals for $\Delta$ for every pairwise entity using 10,000 bootstrapped samples. Lashley et al. (2018) recommend at least 100 detections of a targeted species were collected before estimating activity pattern from camera trap data; in this study, all species detections exceeded the detection threshold (see Results).

**Spatiotemporal Interactions**

Following Karanth et al. (2017), we used multi-response permutation procedures to assess spatiotemporal segregation between cattle and each wild ungulate, which is conditional on the observed space use and temporal activity patterns of the focal species. At camera sites where both cattle and each ungulate co-occurred, we calculated the minimum cattle encounter time for each ungulate and then generated expected statistical distributions of times-to-encounter by randomly assigning encounter times to camera-trap locations in 1000 simulations. We compared the median observed time-to-encounter with a random simulated expected distribution; a larger observed time-to-encounter than expected (assuming species independence) reflects species segregation while a smaller value implies species aggregation. Because cattle moved to the village in winter, spatiotemporal niche analyses were conducted separately in 2013 (from August to November) and in 2014 (from April to July).

## RESULTS

We recorded 1631 detections of wild boar, 3559 detections of roe deer, and 1166 detections of sika deer over 114,854 trapping-days from August 2013 to July 2014. Wild boar and roe deer were photographed at 84% and 92% of the camera stations respectively, while sika deer were only photographed at 40% of the stations (Fig. 1a-c). We observed a total of 3110 cattle encounters with approximately 30% of all stations having cattle presence (Fig. 1d). Tigers ($n = 356$ detections) and leopards ($n = 362$ detections) were photographed at 21.35% and 31.46% of all stations, respectively (Fig. 1e, f).

**Habit Use Modelling**

All covariates were retained because no significant collinearity was detected (VIF < 3.0 and $r$ < 0.7) (Table S2), and including the cattle interactions improved the N-mixture model performance for each species (Table 2). Introducing the cattle encounter rates to the habitat use model (cattle.num) better predicted wild boar and roe deer encounters than the simple presence/absence of cattle (cattle.pres), while the opposite was found for sika deer encounters. As expected, the cattle spatial activity had a significant negative correlation with wild boar and roe deer encounters, while the presence of cattle also correlated negatively with sika deer encounters (Fig. 2 and Table S3).

Wild boars were encountered more often at intermediate elevations (with preference peaking at *ca.* 600 m a.s.l.), farther from roads, closer to settlements, and they preferred valleys and flat slopes (Fig. 2a and Fig. S3). Roe deer were found at higher elevations and farther from roads, and they preferred ridges and avoided sika deer and humans (Fig. 2b and Fig. S4). Sika deer activity was predicted to be at lower elevations and farther from settlements, and they tended to avoid vehicles but preferred flat to moderate slopes, dirt roads and lower forest coverage (Fig. 2c and Fig. S5). In addition, sika deer selected habitats closer to the border, reflecting their recent expansion into China from Russia (also see Fig. 1c). Cattle used lower elevations and valley bottoms and were closer to settlements (Fig. 2d and Fig. S6). They also occurred where there was higher percent tree cover and did not avoid people on foot or vehicles.

The detection probabilities of the three ungulates and domestic cattle were positively correlated with the number of camera days. Roe deer and sika deer were more likely to be detected at locations with lower tiger and leopard presence, but wild boar were associated with lower tiger presence and higher leopard presence. Cattle were more likely to be detected at locations with lower tiger activities. Although overdispersion metrics for the top model for wild boar (*c-hat* =1.16), roe deer (*c-hat* =1.06), sika deer (*c-hat* =1.83) and cattle (*c-hat* =1.26) suggested a slight high-dispersion, we did not find strong spatial patterns in the site-sum randomized-quantile residuals of the N-mixture model for each species (Fig. S2).

**Spatial Co-occurrence**

Wild boar and roe deer did not show substantial spatial overlap or segregation patterns with cattle (*SIF* = 1.07 ± 0.04 SE and 0.99 ± 0.03 SE, respectively, Table 3). However, cattle and sika deer exhibited lower levels of co-occurrence in habitat use (i.e., apparent spatial avoidance, *SIF* = 0.87 ± 0.05 SE) with sika deer occupancy highest at sites where cattle were not detected (*psi*Ba = 0.41 ± 0.03 SE) compared to where they were detected (*psi*BA = 0.34 ± 0.01 SE).

**Daily Activity Patterns**

Cattle activity overlapped with all ungulate species at high rates that ranged from 0.76 (roe deer) to 0.82 (sika deer) (Fig. 3).

**Spatiotemporal Interactions**

When spatiotemporal overlap occurred, we examined the times-to-encounter between the three ungulates and cattle to test for behavioural avoidance. The proportion of independent events recorded in the sites where cattle were absent exceeded 70% among three ungulates (Table S4), and sika deer and cattle co-occurred at the fewest camera sites. The median observed minimum time-to-encounter (ranging from 6.86 to 16.94 days) was significantly greater than the randomly generated time-to-encounter

(ranging from 4.10 to 6.72 days) in both years, suggesting fine-scale behavioural avoidance (Fig. 4 and Table S4).

## DISCUSSION

### Effect of Free-ranging Livestock on Ungulate Species

Our study combined large-scale camera-trap data and multiple spatiotemporal methods to assess the drivers of fine-scale spatiotemporal variation in habitat use by three ungulate species (wild boar, roe deer and sika deer) along the China-Russia border. Our results revealed decreased habitat use or spatial avoidance by all species studied in response to cattle grazing, which supports our hypothesis that cattle grazing have a negative effect on sympatric medium and large-sized herbivores.

Not surprisingly, habitat use by the three ungulates was influenced by topography, humans and land management practices (i.e., grazing). The N-mixture models revealed some separation by all ungulates along the elevation gradient with sika deer responding positively to lower elevations, followed by wild boar, which selected intermediate elevations, and roe deer, which selected higher elevations. The three ungulates also exhibited different responses to TPI. Thus, topographic features may reduce resource competition and promote coexistence among sympatric ungulates in our study area. Due to their tendency to be active during the daytime, the three ungulates avoided roads, vehicles and other areas with people. Elsewhere in Asia, ungulates have been documented to exhibit similar behavioural responses when inhabiting areas with anthropogenic disturbance. In the Russian Far East, the three ungulates strongly avoided areas with high road densities, and sika deer were found far from settlements (Hebblewhite et al. 2014). The relative abundance of wild ungulates declined with the number of villages in the vicinity and increased with the distance to the nearest village in the Himalayan mid-hill landscapes of Nepal (Paudel and Kindlmann 2012). We noted that wild boar thrive near settlements, which likely reflects their preference for agricultural lands along the edges of human developments (Apollonio et al. 2010).

At the landscape scale, our results demonstrate the importance of biotic interactions in shaping distribution patterns and potential range limits. Although species distribution modelling is widely applied in conservation (Romero et al. 2016), most studies exclude species interactions (Wisz et al. 2013), so our results contribute to the low but growing recognition of the influence of biotic interaction on distribution patterns. For example, the marked negative influence of sika deer on the use of camera sites by roe deer suggested that sika deer might cause roe deer to move into areas that they do not use (e.g., higher elevations, Fig. 2), perhaps leading to ecological niche differentiation. This was consistent with a study by Aramilev (2009) which found sika deer occupy roe deer habitats in the Russian Far East and cause roe deer shift to mid-mountain elevations. In particular, the cattle-ungulate interactions provided additional explanatory power and improved model performance for all ungulate species. The best model for each ungulate incorporated cattle interactions (either presence or number of encounters), but ecological differences between wild ungulates resulted in different behavioural responses.

Wild boar and roe deer are highly flexible species that thrive in human-dominated landscapes, and they are now common throughout much of the region (their naïve occupancy was > 80%). Cattle used lower elevations than wild boar and roe deer, and both wild ungulates noticeably reduced their habitat use at low elevations and valley bottoms as more cattle were encountered, suggesting that cattle could compel these two wild species to shift to higher elevations. This was consistent with the findings of Stewart et al. (2002), who demonstrated substantial resource partitioning in the elevations used by elk (*Cervus elaphus nelsoni*), mule deer (*Odocoileus hemionus*) and cattle. In brief, wild boar and roe deer physically distance themselves from cattle herds but do not abandon the habitat at a fine scale; they apparently exhibited fine-scale behavioural avoidance when spatiotemporal overlap occurred at camera sites (see Fig. 4). Similarly, Madhusudan (2004) reported that wild boar, a non-ruminant generalist, did not strongly respond to livestock activities in a tropical Indian wildlife reserve.

For sika deer, the best model included the presence/absence of cattle instead of livestock encounter rates, so as expected, sika deer may be less tolerant of disturbance from livestock than the other two ungulates. The existing levels of grazing by livestock could be high enough to alter the habitat preference of sika deer, irrespective of the intensity. The two-species occupancy model further validated this idea that cattle occupy the resources and limit sika deer dispersal to the west of the border (spatial exclusion, see Fig. 1 and Table 3). Such large-scale competitive exclusion could mean an effective reduction in the extent of suitable habitat available to sika deer. Cattle and sika deer mainly feeding by grazing, roe deer are browsers, while wild boar feed through browsing, grazing or rooting (Ballari and Barrios-Garcia 2014). Food availability and energy requirements could influence their diet and further impact their interactions.

This evidence of negative interactions among cattle and ungulates supports research showing that livestock may be displacing large ungulates, particularly grazing ruminants, or altering their niches in areas of overlap (Madhusudan 2004, Hibert et al. 2010, Dave and Jhala 2011). Long-term livestock grazing lowered chital density by 62% compared to livestock-free areas in the Gir Forest of India (Dave and Jhala 2011). Similar avoidance patterns were also observed in landscapes in the northwestern United States, where elk were displaced by the presence of cattle (Stewart et al. 2002). In summary, we revealed divergent responses of the three ungulates to livestock activities. Madhusudan (2004) suggested that feeding ecology and digestive strategies could play an important role in determining livestock impacts on wild herbivores. In our study area, additional work regarding the diet and foraging behaviour of domestic and wild herbivores is needed to improve our understanding of their co-occurrence relationships.

The data observed in this study offer little support for behaviour-mediated segregation between wild ungulates and livestock. Our results showed a high overall overlap in activity between cattle and all the wild ungulates ($\Delta > 75\%$, Fig. 3), suggesting that the temporal partitioning was not a proximate behavioural response to the presence of livestock.

**Management Implications and Recommendations**

Our results suggest that the presence of cattle and the associated land management may be impeding the recovery of wild ungulate populations in Northeast China. The results of this study have conservation implications in terms of assessing the cascading effects of cattle grazing through a multispecies perspective. Our recent work (Wang et al. 2017, Wang et al. 2018) suggests that long-term livestock husbandry practices may be one of main determinants of tiger and leopard range contractions due to unsustainable pressures on the forest year-round. Moreover, we have speculated that competition between livestock and wild ungulate is a major constraint on the population growth of the two predators and thus have advocated strict grazing controls. Here, we provide further evidence of the negative influences of domestic cattle on three wild prey species, particularly sika deer, and demonstrate the importance of understanding the mechanisms underlying these predator-prey dynamics. In the future, studies of predator–prey dynamics should account for the costs of additional risks caused by indirect effects (i.e., cattle–predator–prey dynamics), as suggested in this study.

Tigers and leopards are now showing a trend towards expanding their range into China (Wang et al. 2015, Dou et al. 2016), and solving the free-ranging livestock problem should be key priority in the new era of conservation. Thus, we suggest that the local government implement policies related to progressively controlling cattle for recovering wild ungulates while simultaneously addressing the economic needs of local communities to ensure the long-term success of tiger and leopard conservation.

A conservative intervention in our study area might be to convert free-ranging livestock to stall feeding, which could reduce the impacts on the forest and conflicts between wildlife and humans. The more progressive intervention would be to ban livestock and redirect ranchers to an ecosystem service project. If it is not feasible to ban all free-range livestock grazing in the study area, we suggest only allowing cattle to enter the forest after the birth peak in spring and early summer and to move cattle to stalls in the village at night. In addition, we strongly encourage local residents to remove wire and electric fence from natural areas to facilitate wildlife movement. These interventions to reduce livestock grazing may rapidly benefit wild herbivores that have been competitively suppressed, as has been observed in India (Madhusudan 2004). The above actions would require better collaboration among different government departments to effectively implement the policy, the establishment of a corresponding monitoring and evaluation system, and a functional law enforcement regime to facilitate the protection of the landscapes that wild ungulate as well as their predators inhabit (Johnson et al. 2016).

## ACKNOWLEDGEMENTS

We thank Daniel Eacker and the anonymous referees for their comments and suggestions that greatly improved our manuscript. This work was supported by the National Key Research and Development Program (2016YFC0500106) and grants from the National Natural Science Foundation of China (31270567, 31700469). Our camera trapping protocol was approved by Chinese State Forestry Administration. Non-invasive camera-trapping technology did not involve direct contact with animals. All study was in accordance with the guidelines approved by The American Society of Mammalogists.

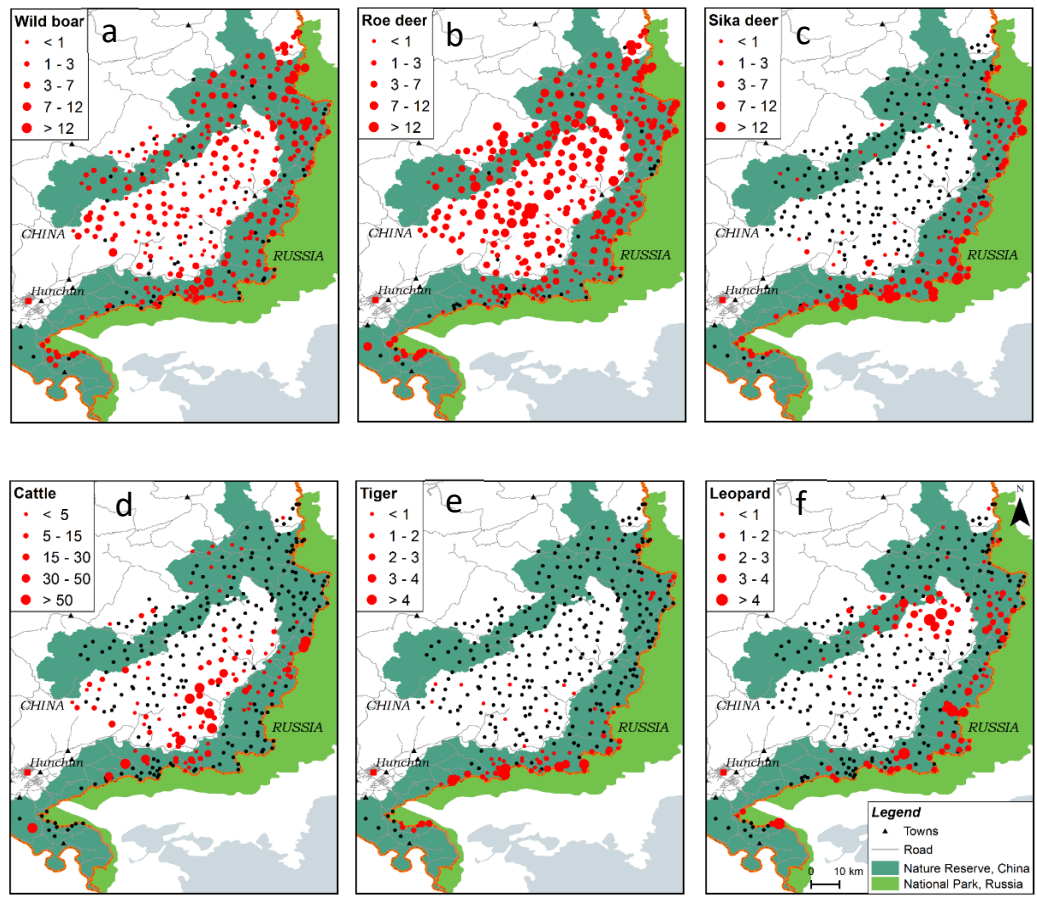

**Figure 1.** Spatial patterns of detection frequency per 100 days for animal species in the camera-trapping study area along the China-Russia border from August 2013 to July 2014: **a** wild boar *Sus scrofa*, **b** roe deer *Capreolus pygarus*, **c** sika deer *Cervus nippon*, **d** cattle *Bos taurus*, **e** Amur tiger *Panthera tigris altaica* and **f** Amur leopard *Panthera pardus orientalis*. Black dots represent sample locations (camera traps) where the species was not observed.

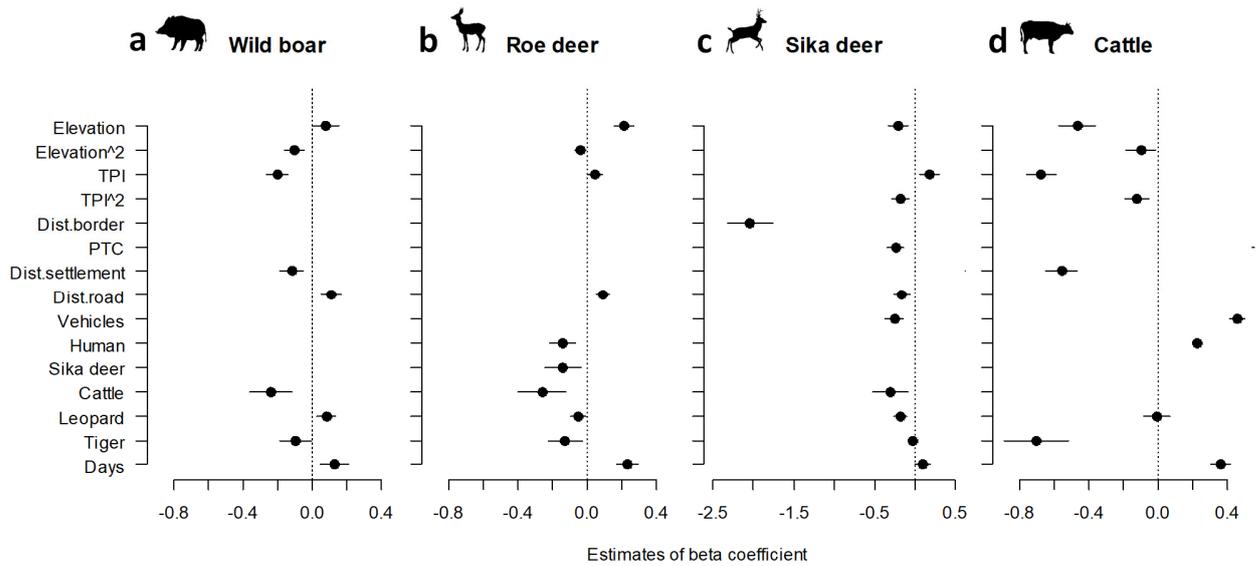

**Figure 2.** Model coefficients (with 95% confidence intervals) of N-mixture models predicting for the influence of abiotic and biotic covariates on the habitat use of the three wild ungulates and domestic cattle along the China-Russia border during August, 2013 to July, 2014. See Table 1 for variable definitions and abbreviations.

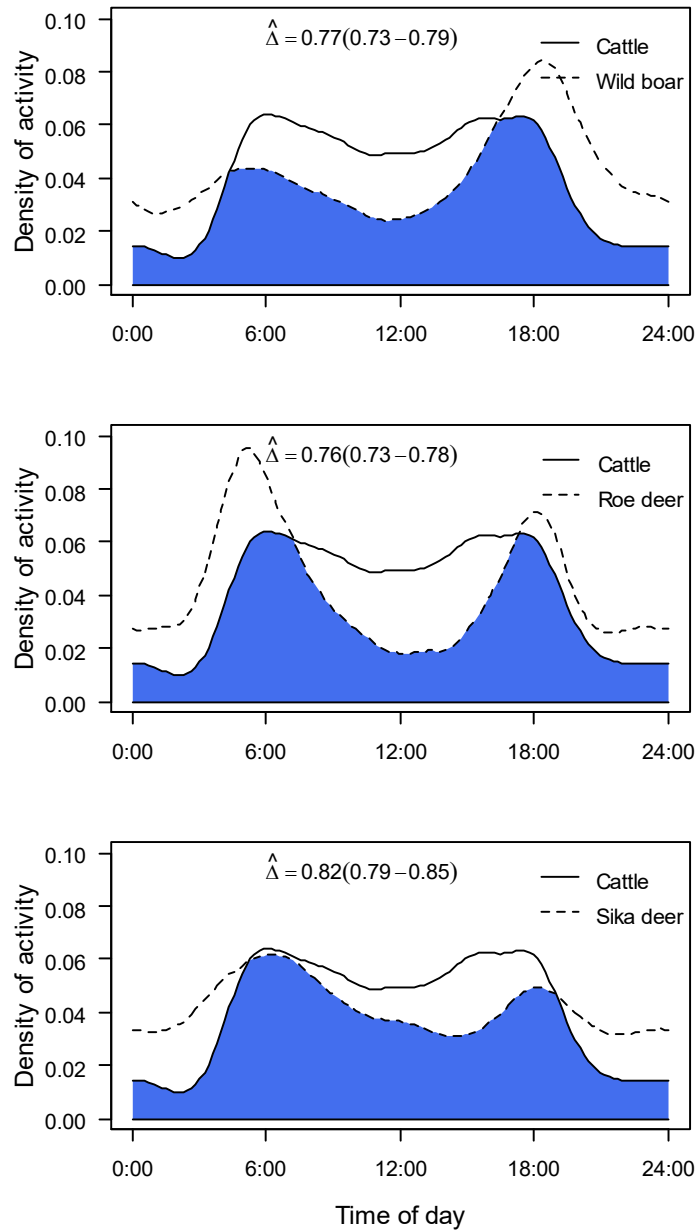

**Figure 3.** Temporal overlap of daily activity patterns between cattle and three wild ungulate species from April to October along the China-Russia border. The estimated overlap is represented by the darkened area and is defined as the area under the curve, which is determined by taking the smaller value of the two activities at each time point.

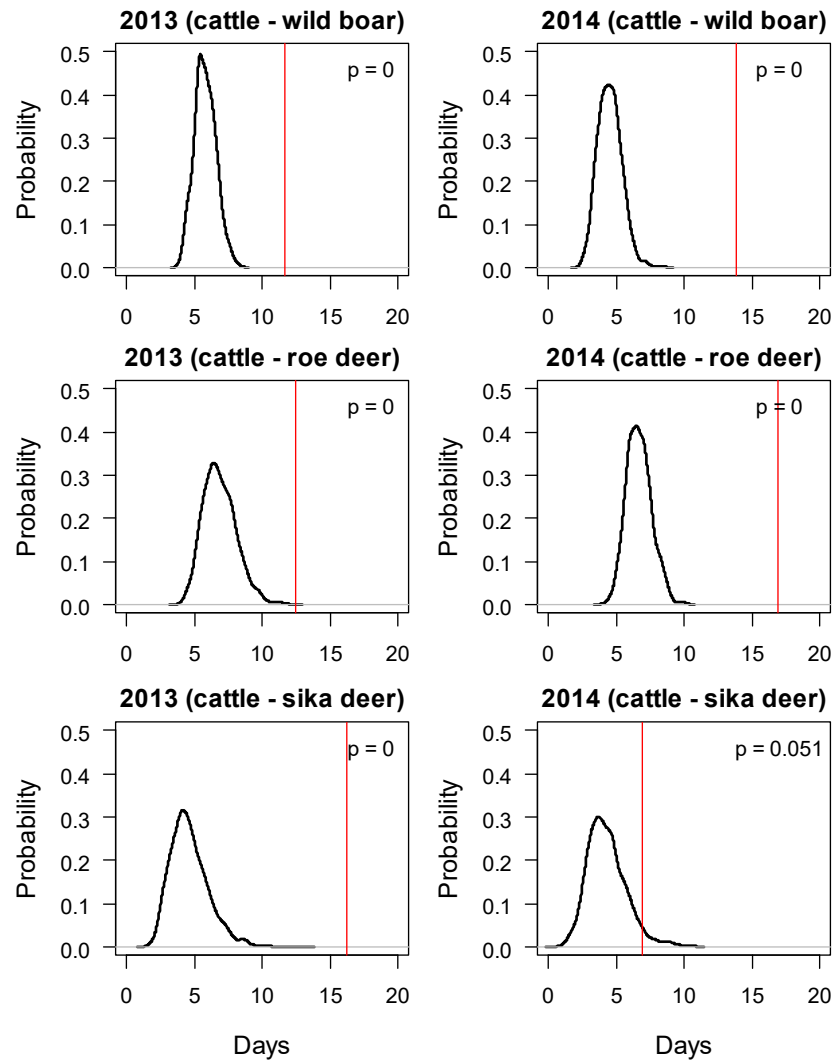

**Figure 4.** Spatiotemporal interactions, as indicated by times-to-encounter, between cattle and wild boar, roe deer and sika deer generated from multi-response permutation procedures along the China-Russia border during August, 2013 to July, 2014. The vertical lines represent the median minimum time-to-encounter between two species, while the area under the curve shows randomly simulated times-to-encounter. The *p*-values, representing the proportions of randomly generated times-to-encounter values that are greater than the observed times-to-encounter, are given for each year.

**Table 1.** Covariates used for N-mixture models to model habitat use by wild boar, roe deer and sika deer along the China-Russia border during August, 2013 to July, 2014.

| Covariate (abbreviation) | Description | Observed range of values | Data source | Component expected to influence |
|---|---|---|---|---|
| Elevation | Numeric (m), elevation of point generated from 30-m DEM | 152-1349 | Shuttle Radar Topography Mission (SRTM) 1 Arc-Second Global [a] | Abundance |
| Topographic position index (TPI) | Numeric, the difference between the elevation of a central pixel and the mean of its surrounding cells. Negative values represent valley bottoms. | -70-109 | Calculated from elevation grids | Abundance |
| Percent tree cover (PTC) | Numeric, percent of pixel that is covered by trees (%) | 22-70 | MODIS/Terra Vegetation Continuous Fields Yearly 250 m (MOD44B) [b] | Abundance |
| Distance to border (Dist.border) | Numeric (km), distance from camera to the nearest border | 0.019-48.21 | Calculated from local geographic information dataset | Abundance |
| Distance to settlement (Dist.settlement) | Numeric (km), distance from camera to the nearest settlement | 0.33-14.89 | Calculated from local geographic information dataset | Abundance |
| Distance to road (Dist.road) | Numeric (km), distance from camera to the nearest road | 0-7.30 | Calculated from local geographic information dataset | Abundance |
| Human presence (Human) | Numeric, encounter rate of people on foot (detections /100 trap-days) | 0-140.48 | Camera trap | Abundance |
| Vehicles | Numeric, encounter rate of vehicles (detections /100 trap-days) | 0-282.97 | Camera trap | Abundance |
| Sika deer | Numeric, encounter rate of sika deer (detections /100 trap-days) | 0-28.92 | Camera trap | Abundance |

| Cattle | Numeric, encounter rate of cattle (detections /100 trap-days) | 0-258.33 | Camera trap | Abundance |
| Cattle presence (Cattle.pres) | Categorical, cattle presence/absence in camera site (0/1) | Indicators of each category (1 or 0) | Camera trap | Abundance |
| Tiger | Numeric, encounter rate of tigers (detections /100 trap-days) | 0-10.72 | Camera trap | Detection |
| Leopard | Numeric, encounter rate of leopards (detections /100 trap-days) | 0-5.29 | Camera trap | Detection |
| Days | Numeric, total days each camera was in operation | 0-365 | Camera trap | Detection |

[a] SRTM dataset (https://lta.cr.usgs.gov/SRTM1Arc)

[b] MODIS vegetation continuous cover/fields (https://lpdaac.usgs.gov/dataset_discovery/modis/modis_products_table/mod44b_v006)

**Table 2.** N-mixture model for interactions of sika deer, roe deer and wild boar with livestock along the China-Russia border during August, 2013 to July, 2014. All models use the same detection covariates (time + days + tiger + leopard). See Table 1 for variable definitions and abbreviations.

|  | Wild boar | | | | Roe deer | | | | Sika deer | | | |
| --- | --- | --- | --- | --- | --- | --- | --- | --- | --- | --- | --- | --- |
|  | Cattle.num | ENV | Cattle.pres | Intercept-only | Cattle.num | Cattle.pres | ENV | Intercept-only | Cattle.pres | Cattle.num | ENV | Intercept-only |
| Elevation | ● | ● | ● |  | ● | ● | ● |  | ● | ● | ● |  |
| Elevation$^2$ | ● | ● | ● |  | ● | ● | ● |  |  |  |  |  |
| TPI | ● | ● | ● |  | ● | ● | ● |  | ● | ● | ● |  |
| TPI$^2$ |  |  |  |  |  |  |  |  | ● | ● | ● |  |
| PTC |  |  |  |  |  |  |  |  | ● | ● | ● |  |
| Dist.settlement | ● | ● | ● |  |  |  |  |  | ● | ● | ● |  |
| Dist.road | ● | ● | ● |  | ● | ● | ● |  | ● | ● | ● |  |
| Dist.border |  |  |  |  |  |  |  |  | ● | ● | ● |  |
| Human |  |  |  |  | ● | ● | ● |  |  |  |  |  |
| Vehicles |  |  |  |  |  |  |  |  | ● | ● | ● |  |
| Sika deer |  |  |  |  | ● | ● |  |  |  |  |  |  |
| Cattle | ● |  |  |  | ● |  |  |  |  | ● |  |  |
| Cattle.pres |  |  | ● |  |  | ● |  |  | ● |  |  |  |
| $K$ | 12 | 11 | 12 | 2 | 13 | 13 | 11 | 2 | 15 | 15 | 14 | 2 |
| AIC | 7706.76 | 7724.40 | 7725.65 | 7947.99 | 12365.00 | 12374.03 | 12384.00 | 12951.93 | 4555.37 | 4558.60 | 4560.90 | 5772.54 |
| ΔAIC | 0 | 17.65 | 18.88 | 241.23 | 0 | 9.02 | 18.51 | 586.91 | 0 | 3.23 | 5.55 | 1217.17 |
| $w_i$ | 1 | 0 | 0 | 0 | 0.99 | 0.01 | 0 | 0 | 0.79 | 0.16 | 0.05 | 0 |

*Notes:* $K$ is the number of parameters; ΔAIC is the difference in AIC relative to the best model; and $w_i$ is the Akaike weight that indicates the relative support for each model. Black points represent the covariates included in the corresponding model.

**Table 3.** Estimates of parameters for the two-species occupancy model for cattle (species A) and three wild ungulates (species B). Camera-trapping data were collected from August 2013 to July 2014 by the long-term Tiger and Leopard Observation Network (TLON) from 356 camera sites along the China-Russia border.

| Parameters | Cattle - Wild boar | Cattle - Roe deer | Cattle - Sika deer |
|---|---|---|---|
| $psi$A | 0.30 (0.02) | 0.30 (0.02) | 0.30 (0.02) |
| $psi$BA | 0.93 (0.04) | 0.88 (0.03) | 0.34 (0.05) |
| $psi$Ba | 0.84 (0.03) | 0.90 (0.02) | 0.41 (0.03) |
| $p$A | 0.34 (0.01) | 0.34 (0.01) | 0.34 (0.01) |
| $p$B | 0.20 (0.01) | 0.33 (0.01) | 0.24 (0.01) |
| $r$B | 0.19 (0.01) | 0.27 (0.01) | 0.18 (0.02) |
| SIF | 1.07 (0.04) | 0.99 (0.03) | 0.87 (0.05) |

# Supplemental Information

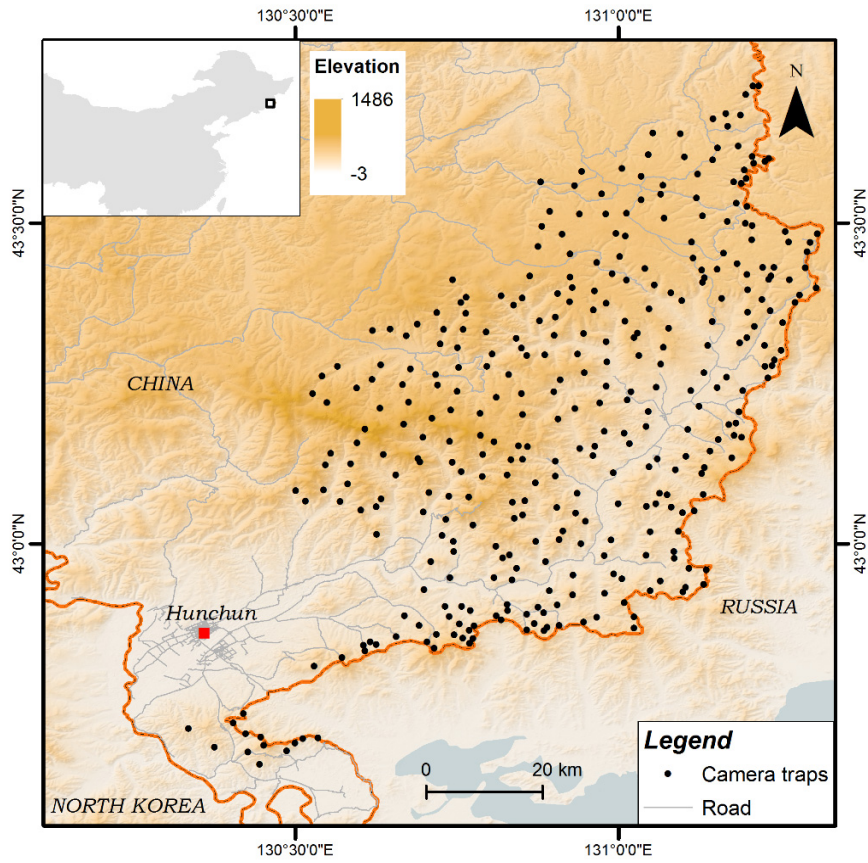

**Figure S1.** Location of the camera-trapping study area along the China-Russia border. Camera trap was conducted during August, 2013 to July, 2014.

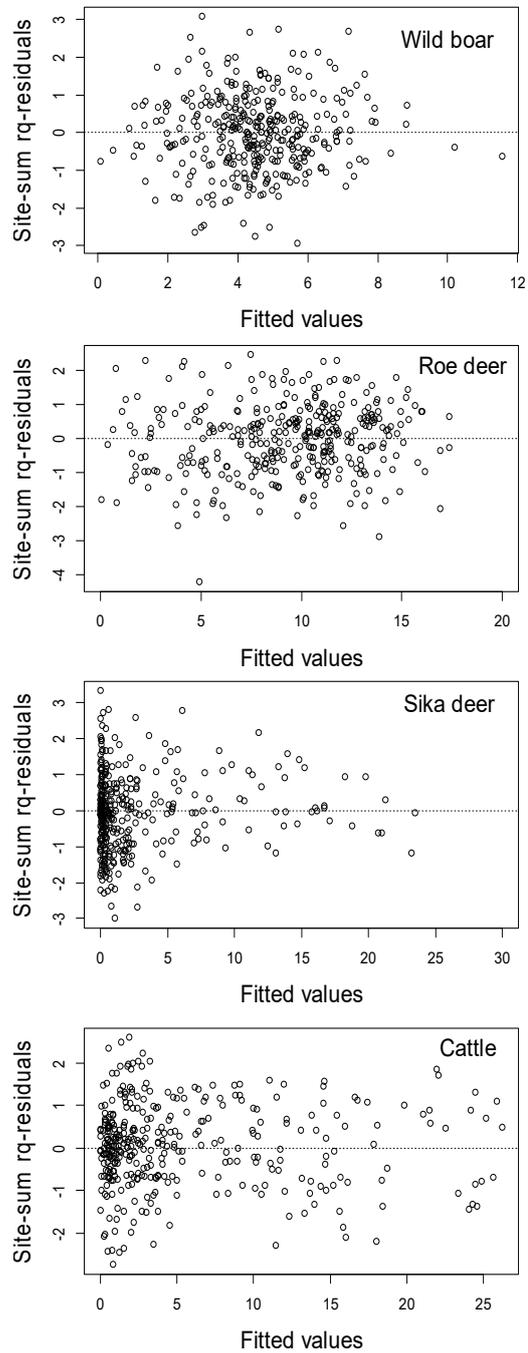

**Figure S2.** Site-sum randomized-quantile residuals against fitted values for fits of N-mixture models to the wild boar, roe deer, sika deer and cattle data.

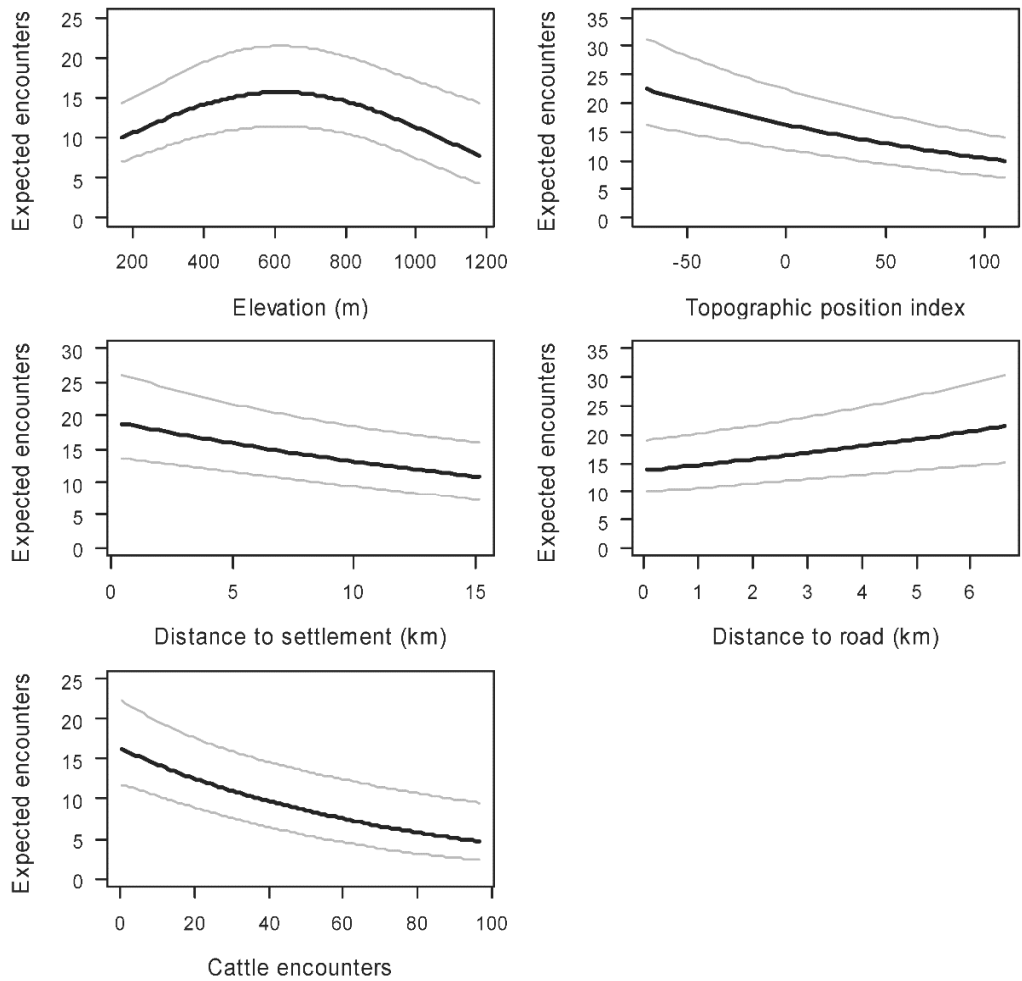

**Figure S3.** Wild boar predicted encounters as a function of different covariates at a given sampling occasion (month) with 95% confidence intervals along the China-Russia border during August, 2013 to July, 2014.

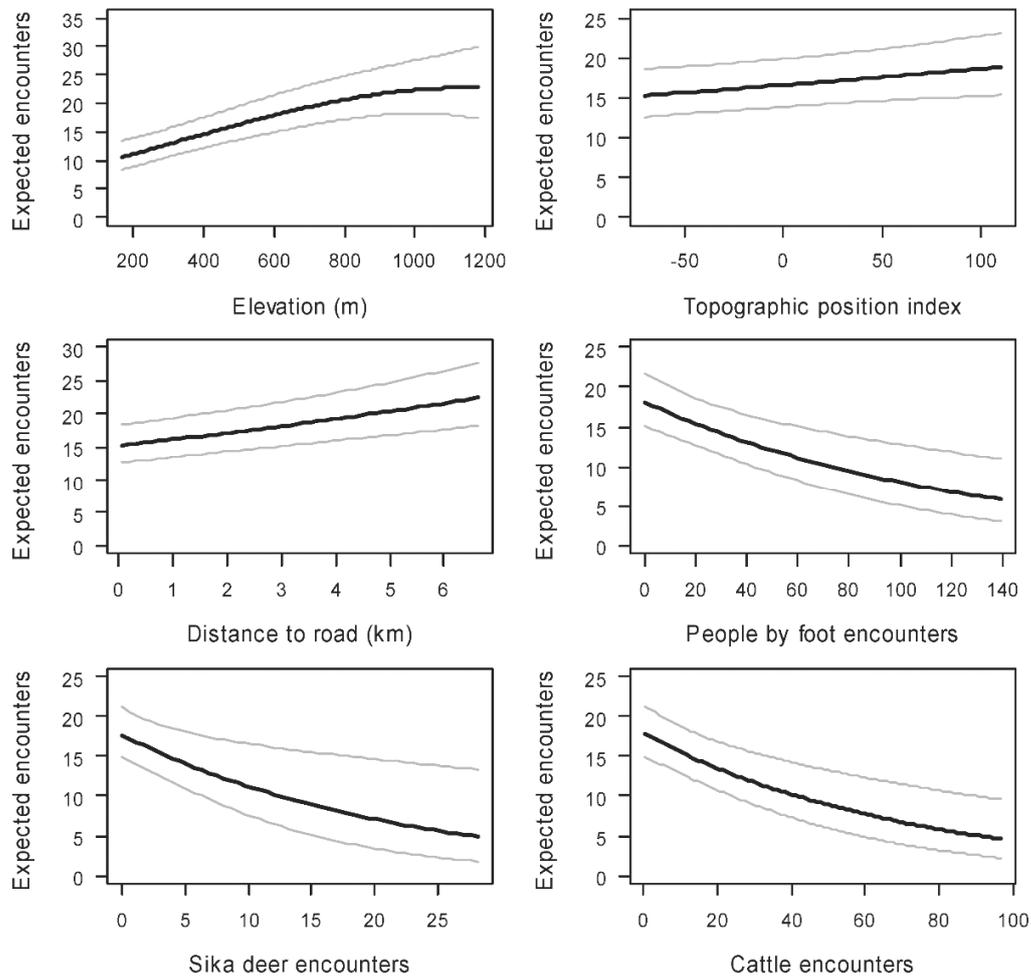

**Figure S4.** Roe deer predicted encounters as a function of different covariates at a given sampling occasion (month) with 95% confidence intervals along the China-Russia border during August, 2013 to July, 2014.

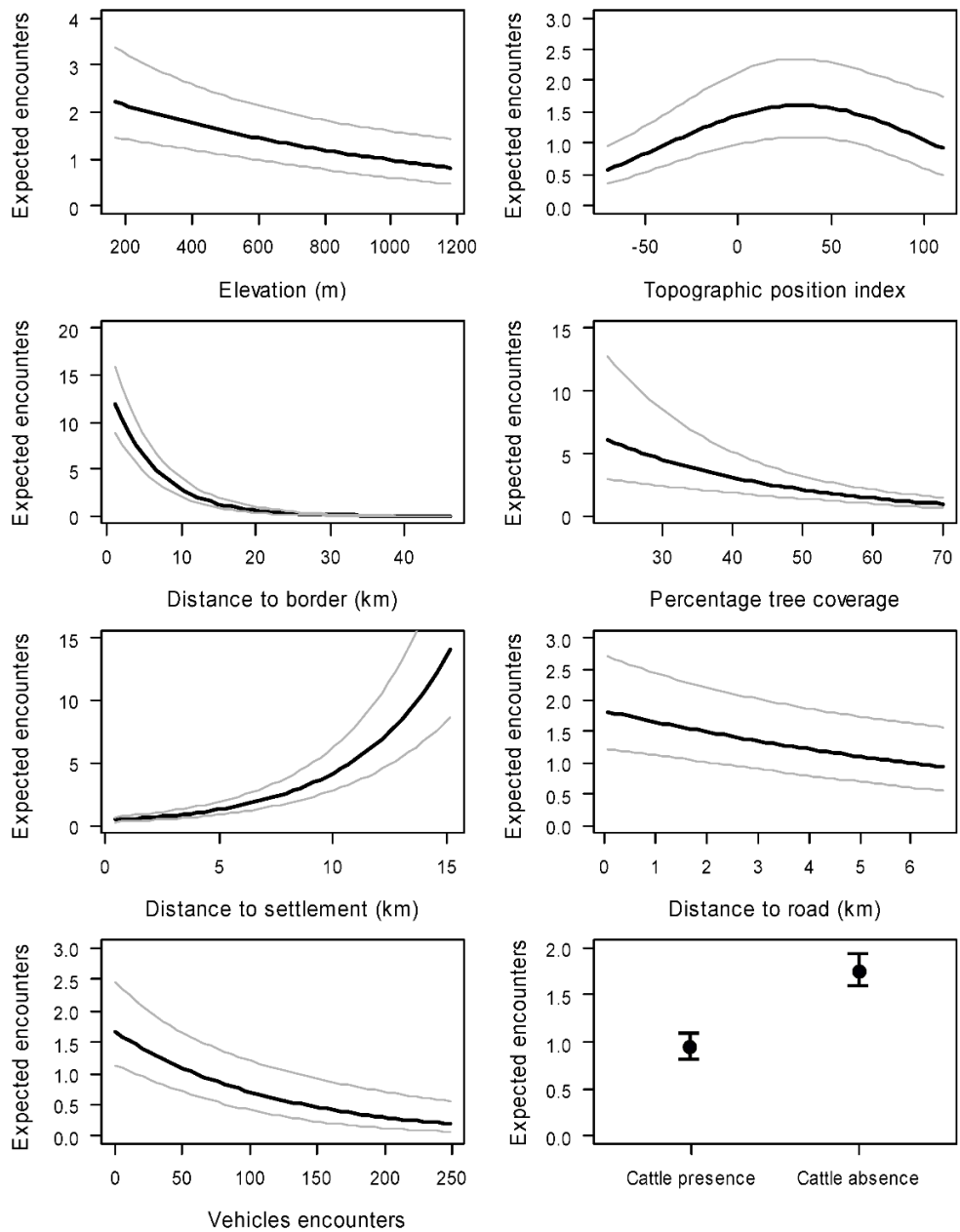

**Figure S5.** Sika deer predicted encounters as a function of different covariates at a given sampling occasion (month) with 95% confidence intervals along the China-Russia border during August, 2013 to July, 2014.

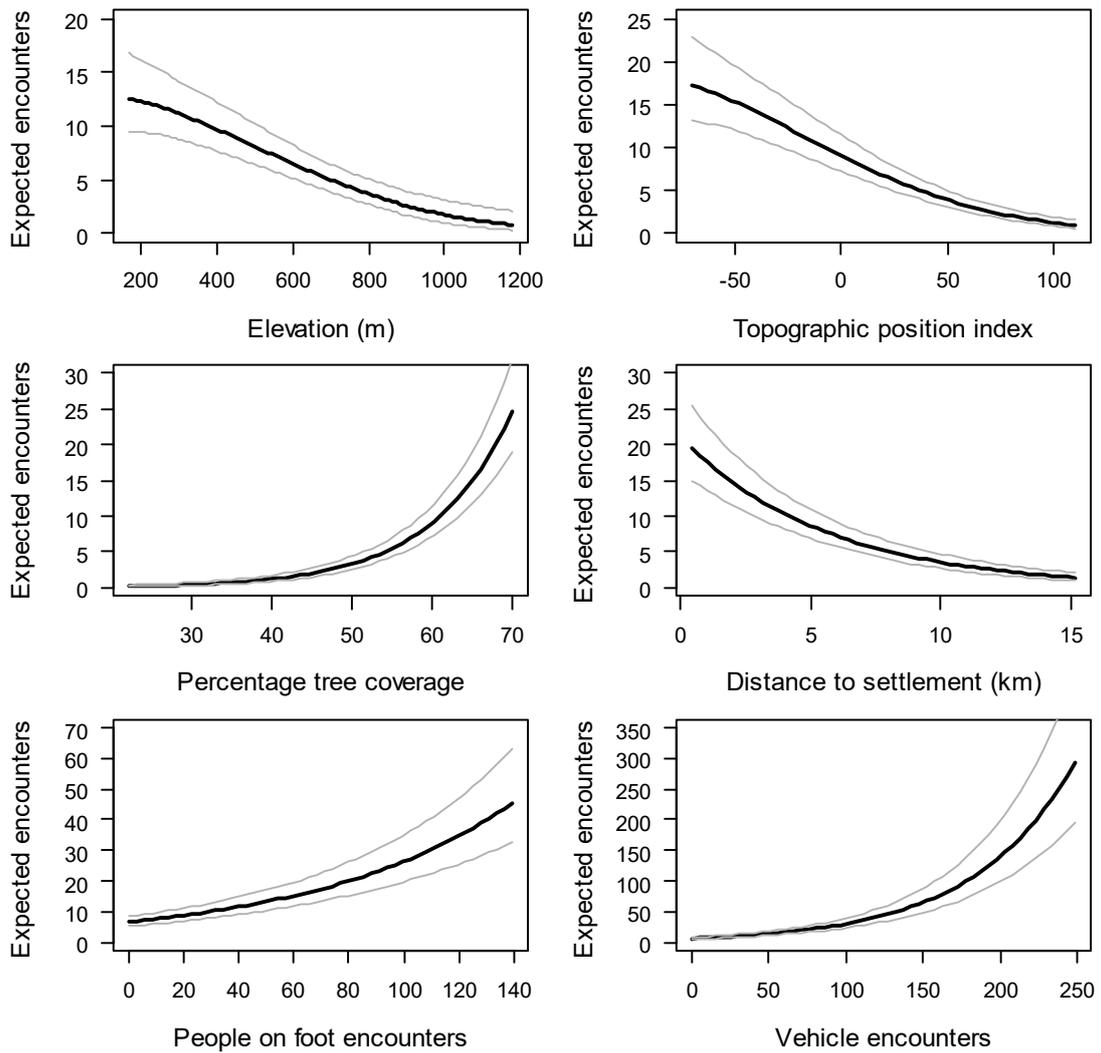

**Figure S6.** Cattle predicted encounters as a function of different covariates at a given sampling occasion (month) with 95% confidence intervals along the China-Russia border during August, 2013 to July, 2014.

**Table S1** Parameters used in the conditional two-species occupancy model for cattle (dominant species A) and the three ungulate species (subordinate species B); table adapted from Richmond *et al.* (2010). We collected the data for this analysis from 2013 to 2014 at 356 sites along the China-Russia border.

| Parameter | Description |
|---|---|
| $\psi$A | Probability of occupancy for species A |
| $\psi$BA | Probability of occupancy for species B, given species A is present |
| $\psi$Ba | Probability of occupancy for species B, given species A is absent |
| $p$A | Probability of detection for species A, given species B is absent |
| $p$B | Probability of detection for species B, given species A is absent |
| $r$A | Probability of detection for species A, given both species are present |
| $r$B | Probability of detection for species B, given both species are present |
| SIF | Species interaction factor, an *SIF* of 1.0 indicates no interaction (e.g. species use space independent of one another), while an *SIF* > 1.0 indicates co-occurrence (e.g. occur together more often than expected if independent) and an *SIF* < 1.0 indicate avoidance (e.g. occur together less often than expected if independent) |

**Table S2.** Pearson's correlation coefficients between the covariates used for N-mixture models of wild boar, roe deer, sika deer and cattle along the China-Russia border during August, 2013 to July, 2014. See Table 1 for variable definitions and abbreviations.

|  | Elevation | TPI | PTC | Dist.settlement | Dist.road | Dist.border | Human | Vehicles | Sika deer | Cattle |
|---|---|---|---|---|---|---|---|---|---|---|
| TPI | 0.29 | | | | | | | | | |
| PTC | 0.37 | 0.02 | | | | | | | | |
| Dist.settlement | 0.46 | -0.07 | 0.27 | | | | | | | |
| Dist.road | -0.01 | 0.08 | 0.08 | 0.05 | | | | | | |
| Dist.border | 0.63 | 0.17 | 0.25 | 0.40 | 0.02 | | | | | |
| Human | -0.31 | -0.33 | -0.17 | -0.01 | -0.10 | -0.26 | | | | |
| Vehicles | -0.14 | -0.20 | -0.20 | 0.01 | -0.12 | -0.17 | 0.63 | | | |
| Sika deer | -0.25 | -0.10 | -0.05 | 0.12 | -0.09 | -0.34 | 0.10 | 0.02 | | |
| Cattle | -0.20 | -0.14 | -0.12 | -0.12 | -0.01 | -0.08 | 0.47 | 0.28 | -0.05 | |
| Cattle.pres | -0.28 | -0.09 | -0.18 | -0.16 | -0.01 | -0.04 | 0.12 | 0.04 | -0.12 | 0.32 |

Table S3. The best supported zero-inflated Poisson N-mixture models explaining the relative abundance of wild boar, roe deer, sika deer and cattle as indicated by the parameter estimates, standard errors (SE) and *p* value along the China-Russia border during August, 2013 to July, 2014. "-" represent statistically dot not correlate covariate for corresponding species. See Table 1 for variable definitions and abbreviations.

| Covariate | Wild boar | | | Roe deer | | | Sika deer | | | Cattle | | |
| --- | --- | --- | --- | --- | --- | --- | --- | --- | --- | --- | --- | --- |
| | Estimate | SE | *p* value | Estimate | SE | *p* value | Estimate | SE | *p* value | Estimate | SE | *p* value |
| **Habitat use** | | | | | | | | | | | | |
| Elevation | 0.081 | 0.039 | 0.0401 | 0.214 | 0.029 | 0.0000 | -0.211 | 0.064 | 0.0009 | -0.466 | 0.054 | 0.0000 |
| Elevation $^2$ | -0.103 | 0.030 | 0.0006 | -0.037 | 0.016 | 0.0196 | | | 0.0000 | -0.098 | 0.043 | 0.0245 |
| TPI | -0.201 | 0.031 | 0.0000 | 0.046 | 0.022 | 0.0385 | 0.176 | 0.063 | 0.0051 | -0.673 | 0.044 | 0.0000 |
| TPI $^2$ | - | - | - | - | - | - | -0.186 | 0.056 | 0.0008 | -0.123 | 0.037 | 0.0008 |
| Dist.border | - | - | - | - | - | - | -2.042 | 0.145 | 0.0000 | - | - | - |
| PTC | - | - | - | - | - | - | -0.240 | 0.053 | | 0.627 | 0.042 | 0.0000 |
| Dist.settlement | -0.118 | 0.034 | 0.0005 | - | - | - | 0.736 | 0.057 | 0.0000 | -0.556 | 0.048 | 0.0000 |
| Dist.road | 0.110 | 0.028 | 0.0001 | 0.091 | 0.020 | 0.0000 | -0.166 | 0.055 | 0.0024 | - | - | - |
| Vehicles | - | - | - | - | - | - | -0.257 | 0.061 | 0.0000 | 0.458 | 0.023 | 0.0000 |
| Human | - | - | - | -0.139 | 0.039 | 0.0004 | - | - | - | 0.227 | 0.014 | 0.0000 |
| Sika deer | - | - | - | -0.141 | 0.054 | 0.0097 | - | - | - | - | - | - |
| Cattle | -0.238 | 0.063 | 0.0002 | -0.258 | 0.070 | 0.0002 | -0.306 | 0.115 | 0.0075 | - | - | - |
| **Detection** | | | | | | | | | | | | |
| Leopard | 0.083 | 0.027 | 0.0026 | -0.051 | 0.022 | 0.0205 | -0.179 | 0.042 | 0.0000 | -0.006 | 0.039 | 0.8690 |
| Tiger | -0.095 | 0.046 | 0.0395 | -0.125 | 0.051 | 0.0146 | -0.038 | 0.024 | 0.1100 | -0.700 | 0.095 | 0.0000 |
| Days | 0.128 | 0.043 | 0.0032 | 0.232 | 0.032 | 0.0000 | 0.096 | 0.048 | 0.0458 | 0.362 | 0.029 | 0.0000 |

**Table S4.** Summary of times-to-encounter between cattle and wild boar, roe deer and sika deer. We collected the camera trapping data for this analysis from 2013 to 2014 at 356 sites along the China-Russia border.

|  | Wild boar | | Roe deer | | Sika deer | |
|---|---|---|---|---|---|---|
|  | 2013 | 2014 | 2013 | 2014 | 2013 | 2014 |
| Number of independent events | 872 | 412 | 1147 | 1484 | 399 | 514 |
| Number of events recorded in the sites where cattle were absent | 636 | 320 | 982 | 1229 | 351 | 485 |
| Proportion of events recorded in the sites where cattle were absent | 72.94 | 77.67 | 85.61 | 82.82 | 88 | 94.36 |
| Number of camera sites that observed species co-occurrence with cattle | 73 | 34 | 49 | 60 | 17 | 16 |
| Median observed minimum time-to-encounter (d) | 11.64 | 13.81 | 12.4 | 16.94 | 16.2 | 6.86 |
| Expected median randomly simulated time-to-encounter (d) | 5.74 | 4.45 | 6.72 | 6.58 | 4.42 | 4.10 |
| *p*-value | 0 | 0 | 0 | 0 | 0 | 0.051 |

Note: The *p*-values represent the proportion of randomly generated time-to-encounter values that are greater than the observed time-to-encounter.